\documentclass[aps,pra,twocolumn,superscriptaddress]{revtex4}

\usepackage{amsmath,amssymb}
\usepackage{multirow}
\usepackage{xcolor}
\usepackage{srcltx}
\usepackage{hyperref,graphicx}

\usepackage[utf8]{inputenc}

\def\be{\begin{equation}}
\def\ee{\end{equation}}
\def\beq{\begin{equation}}
\def\eeq{\end{equation}}
\newcommand{\ket}[1]{|#1\rangle}
\newcommand{\bra}[1]{\langle#1|}

\newcommand{\fisicarm}{Dipartimento di Fisica, Sapienza Universit\`{a} di Roma, Piazzale Aldo Moro, 5, I-00185 Roma, Italy}
\newcommand{\ino}{Istituto Nazionale di Ottica, Consiglio Nazionale delle Ricerche (INO-CNR), Largo Enrico Fermi, 6, I-50125 Firenze, Italy}

\newcommand{\ifn}{Istituto di Fotonica e Nanotecnologie, Consiglio Nazionale delle Ricerche (IFN-CNR), Piazza Leonardo da Vinci, 32, I-20133 Milano, Italy}
\newcommand{\fisicami}{Dipartimento di Fisica, Politecnico di Milano, Piazza Leonardo da Vinci, 32, I-20133 Milano, Italy}
\newcommand{\pad}{Department of Information Engineering, University of Padova, I-35131 Padova, Italy}

\begin{document}

\title{Two-particle bosonic-fermionic quantum walk via integrated photonics}

\author{Linda Sansoni}
\affiliation{\fisicarm}

\author{Fabio Sciarrino}
\affiliation{\fisicarm}
\affiliation{\ino}

\author{Giuseppe Vallone}
\affiliation{\fisicarm}
\affiliation{\pad}

\author{Paolo Mataloni}
\affiliation{\fisicarm}
\affiliation{\ino}

\author{Andrea Crespi}
\affiliation{\ifn}
\affiliation{\fisicami}

\author{Roberta Ramponi}
\affiliation{\ifn}
\affiliation{\fisicami}

\author{Roberto Osellame}
\affiliation{\ifn}
\affiliation{\fisicami}

\begin{abstract}
Quantum walk represents one of the most promising resources for the simulation of physical quantum systems, and has also emerged as an alternative to the standard circuit model for quantum computing. 
Here we investigate how the particle statistics, either bosonic or fermionic, influences a two-particle discrete quantum walk. Such experiment has been realized by exploiting polarization entanglement to simulate the bunching-antibunching feature of non interacting bosons and fermions. To this scope a novel three-dimensional geometry for the waveguide circuit is introduced, which allows accurate polarization independent behaviour, maintaining a remarkable control on both phase and balancement.
\end{abstract}

\maketitle

In the framework of quantum information processing, quantum walk has attracted much attention in the last few years \cite{chil09prl}. Quantum walk is an extension of the classical random walk: a walker on a lattice ``jumping'' between different sites with a given probability. The features of the quantum walker are interference and superposition which lead to a non-classical dynamic evolution. Two different cases may be considered: discrete- and continuous-time quantum walks  \cite{kemp03cph}. 
The properties of these two walks have shown several similarities\cite{stra06pra}, however, the discrete quantum walk exhibits a higher flexibility due to the possibility of tailoring the quantum coin properties to investigate different dynamic scenarios \cite{schr10prl,schr11prl}. By endowing the walker with quantum properties, many new interesting effects appear: quantum walks allow the speed-up of search algorithms \cite{poto09pra} and the realization of universal quantum computation \cite{chil09prl}. Moreover it has been recently shown that quantum walks with a large number of sites exhibit a highly nontrivial dynamics, including localization and recurrence \cite{stef08prl,lahi10prl}. Within this scenario, a possible application is the investigation on biophysical systems, like the energy transfer process within photosynthesis \cite{mohs08jcp}. 

Single-particle quantum walks yield an exponential computational gain with respect to classical random walks
; {it can be noted that they have an exact mapping to classical wave phenomena and therefore they can be implemented using purely classical resources. On the other hand, quantum walks of more than one indistinguishable particle can provide an additional computational power that scales exponentially with the resources employed. This could be used to improve simulation performances in complex tasks, e.g. the graph isomorphism problem \cite{gamb10pra}. However, they need quantum resources to be implemented, since classical theory no longer provides a sufficient description.

Different experimental implementations of single particle quantum walks were performed with trapped atoms \cite{kars09sci}, ions \cite{schm09prl,zahr10prl}, energy levels in NMR schemes \cite{ryan05pra}, photons in waveguide structures \cite{pere08prl}, and in a fiber loop configuration \cite{schr10prl,schr11prl}.
Very recently quantum walks of two identical photons have been performed \cite{peru10sci,owen11njp}. However, up to now no experimental demonstration on how the particle statistics, either bosonic or fermionic, influences a two-particle quantum walk has been reported.

In this work, we report on the implementation of a {discrete} quantum walk for entangled particles. By changing the symmetry of entanglement we can simulate the quantum dynamics of the walks of two particles with bosonic or fermionic statistic. These results are made possible by the adoption of novel geometries in integrated optical circuits fabricated by femtosecond laser pulses, which {preserve} the indistinguishability of the two polarizations as well as {provide} high phase accuracy and stability.
In the discrete quantum walk the walker is represented by a quantum particle -such as an electron, atom or photon- with an additional degree of freedom spanning a two-dimensional space and named the ``quantum coin'' (QC).
At any given time the particle may be in a superposition of the two basis states, up ($\ket{U}$) or down {($\ket{D}$),} representing the two ``coin faces'' (Fig. \ref{fig:BSwalk}a).  
The QC state directs the motion of the particle and the stochastic evolution by
an unitary process. A key difference with the classical case is that the many possible paths of the quantum walker may exhibit interference, leading to a very different probability distribution of finding the walker at a given location.

Indeed, the evolution of the walk can be described with the step operator $E=\sum_j{\ket{j-1}\bra{j}\otimes\ket{U}\bra{U}+\ket{j+1}\bra{j}\otimes\ket{D}\bra{D}}$, where $\ket{j-1}\bra{j}$ and $\ket{j+1}\bra{j}$ stand for the operators which move the particle in the higher and lower position of the lattice, respectively. The coherent action of the step operator $E$ and coin tossing leads to entanglement between the position and the internal degree of freedom. After several steps, the counterintuitive profile of the quantum walk probability distribution emerges as a result of quantum interference among multiple paths.

\begin{figure}
\includegraphics[width=8.5cm]{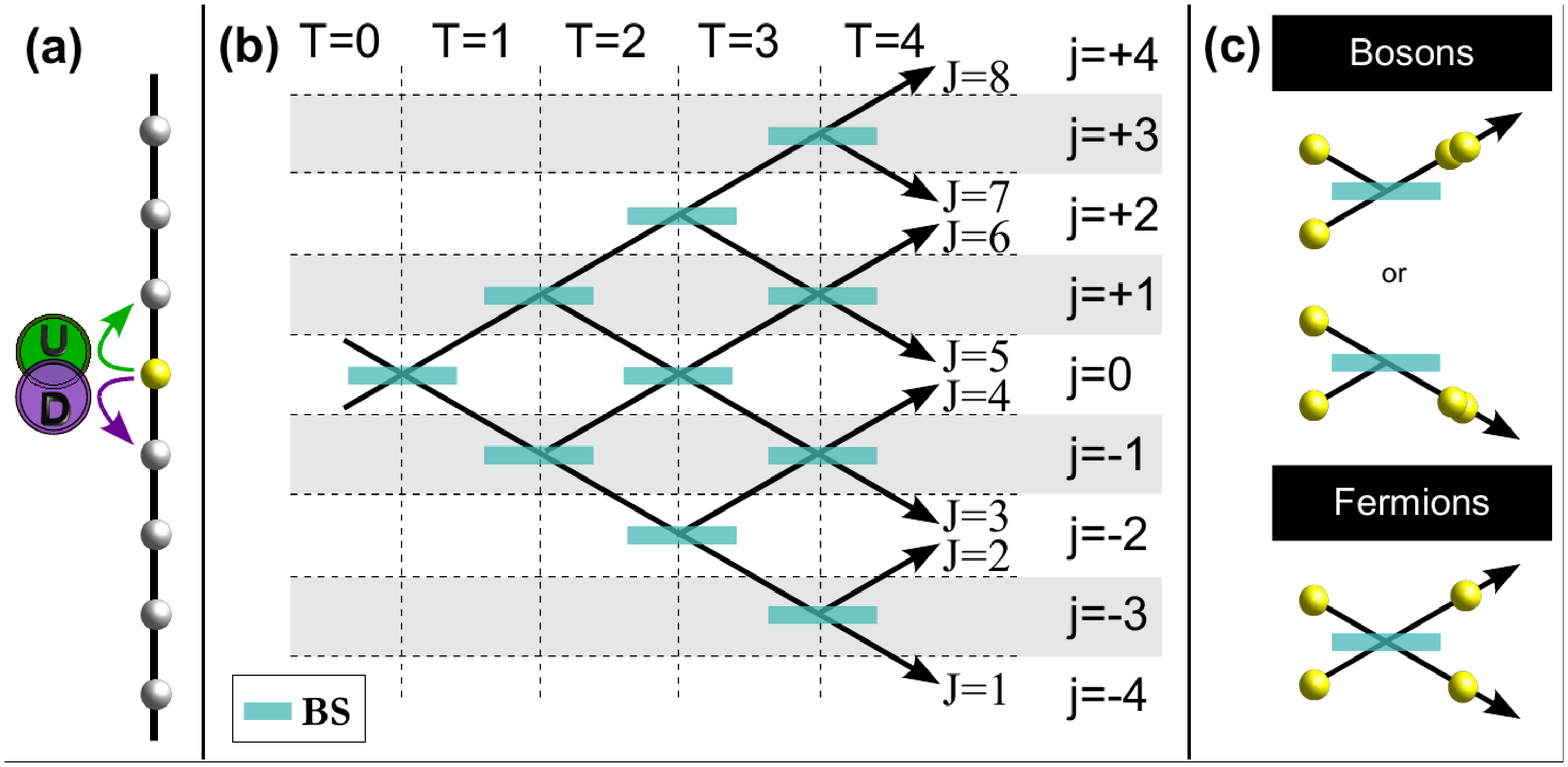}
\caption{(a) Unidimensional quantum walk: depending on the result of the coin toss the walker moves upward (U) or downward (D). (b) Scheme of an array of beam splitters (BSs) for a four-steps quantum walk. Vertical dashed lines indicate the steps $T$ of the quantum walk and the horizontal strips represent the position $\ket{j}$ of the walker. In an array with an even (odd) number of steps the output ports $J$ {are grouped} into {the} even (odd) final positions $\ket{j}$ of the walker. (c) Different behaviours of bosons and fermions on a BS.}
\label{fig:BSwalk}
\end{figure}

More complex distributions arise when two particles are injected into the same quantum walk.
An important perspective is to combine in the same platform quantum walk and entangled states \cite{omar06pra,path07pra}.
In this case both initial and final state of the walkers are entangled states and,
depending on the entanglement symmetry, different final distributions may be observed.
By changing the symmetry of the entangled state we can simulate the quantum walk of two particles with integer or semi-integer spin \cite{SI:prob}. The fermionic and bosonic behaviors drastically influence the dynamics of their quantum walk on the line.

The simulation of single particle quantum walks on a line can be implemented
using single photon states, beam splitters, phase shifters, and photodetectors \cite{jeon04pra,rohd11njp}. 
The quantum dynamics is achieved by an array of balanced beamsplitters (\textit{BS}s) as shown in {Fig. \ref{fig:BSwalk}b}, each vertical line of beam splitters representing a step of the quantum walk. Horizontal strips represent the position $\ket{j}$ of the walker.
If a photon, at time $T$ and in the strip $j$, is incident downward $\ket{D}$ (upward $\ket{U}$) on the BS
we can represent its state as $\ket{j,D}_T$ ($\ket{j,U}_T$). The transition from time $T$ to time $T+1$
 is given by the BS operator $\ket{j,D}_T\rightarrow\frac{1}{\sqrt2}(\ket{j-1,D}_{T+1}-\ket{j+1,U}_{T+1})$,
$\ket{j,U}_T\rightarrow\frac{1}{\sqrt2}(\ket{j+1,U}_{T+1}+\ket{j-1,D}_{T+1})$.
This operation simultaneously implements the coin (precisely the Hadamard
coin $C=\frac{1}{\sqrt2}\left(\begin{smallmatrix}1 &    1\\1&-1\end{smallmatrix}\right)$)
and step {operator} $E$. Note that, if the particle starts at position
$\ket{j=0}$, at even (odd) times it will occupy only even (odd) positions.

Provided that all the optical devices used in the walk are polarization insensitive, the polarization degree of freedom may 
be exploited to {entangle the photons injected} into the \textit{BS} arrays. Moreover by changing the entangled state from 
a symmetric one, such as the triplet, into an antisymmetric one, the singlet, it is possible to mimic the quantum dynamics 
of two non-interacting bosonic and fermionic particles (see {Fig. \ref{fig:BSwalk}c}). It must be noticed that the experimental realization of such a network of \textit{BS}s is exceedingly 
difficult with bulk optics, even for a small number of steps, since it requires a quadratically growing number of elements. 
Furthermore, for a correct operation of the quantum walk, the 
phase introduced by the optical paths, passing from each beam splitter to the following, must be controlled and stable.\\
Our approach exploits an integrated waveguide architecture, which allows to concentrate a large number of optical elements on a small chip and 
to achieve intrinsic phase stability due to the monolithic structure. In a waveguide implementation \textit{BS}s are replaced by directional 
couplers (\textit{DC}s) i.e. structures in which two {waveguides, brought close together for a certain interaction length,} couple by evanescent field. 
\begin{figure}
	\includegraphics[width=8.5cm]{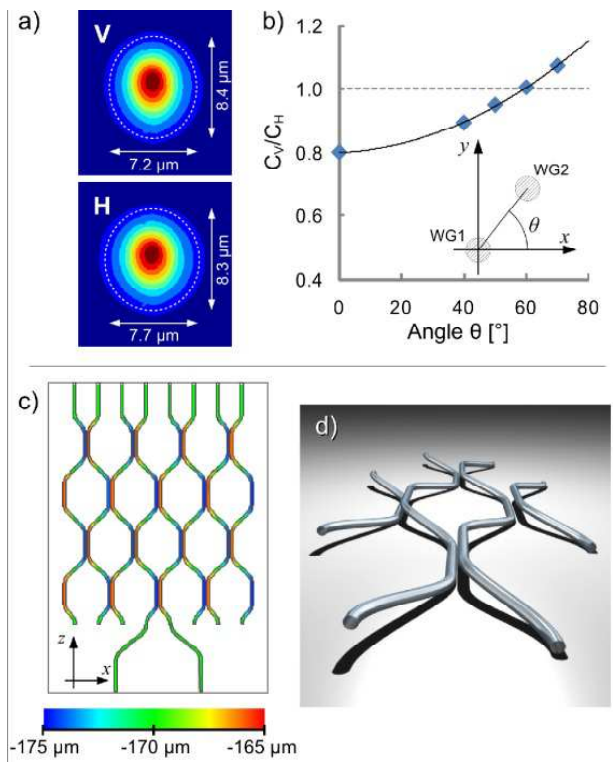}
	\caption{
		\textbf{Integrated optical circuits.} \textit{a)} Measured intensity profile for the guided modes with polarization V and H, at $806 nm$ wavelength. The $1/e^2$ dimensions are reported. \textit{b)} Ratio of the estimated coupling coefficient for polarization V ($C_V)$ and polarization H ($C_H$) in directional couplers fabricated with different angles $\theta$ between the waveguides (see inset), but fixed interaction length ($3 mm$) and distance ($11 \mu m$). The fitting line is a guide to the eye. \textit{c)} Schematic of the network of directional couplers fabricated for implementing a 4 steps quantum walk. The color coding indicates the writing depth of the waveguides, which is varying from point to point. \textit{d)} 3D representation of the basic cell of the network, which acts as a Mach-Zehnder interferometer}
		\label{fig:couplers}
\end{figure}

To realize the integrated optical circuits we adopted the femtosecond laser writing technology \cite{gatt08npo, dval09joa}. 
Briefly, nonlinear absorption of focused femtosecond pulses is exploited to induce permanent and localized refractive index increase in 
transparent materials. Waveguides are directly fabricated in the bulk of the substrate by translation of the sample at constant velocity with 
respect to the laser beam, along the desired path. Since it is a single-step and maskless process, this technique allows rapid and cost-effective 
prototyping of new devices. Furthermore, it has intrinsic three-dimensional possibilities which have indeed been exploited in this work.

\begin{figure}[ht]
\includegraphics[width=8.5cm]{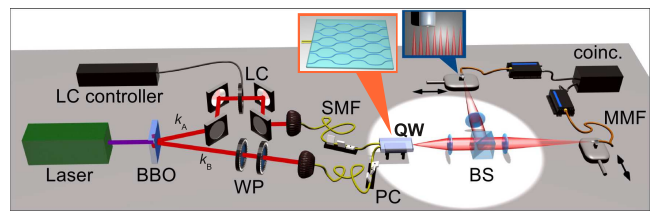}
\caption{ Experimental setup can be divided into three parts.
(i) The source: polarization entangled photon pairs at wavelength $\lambda=806nm$ generated via
spontaneous parametric down conversion in a 1.5mm $\beta$-barium borate  
crystal (BBO) cut for type-II non-collinear phase matching, 
pumped by a CW diode laser with power $P = 50mW$ \cite{kwia95prl}.
Waveplates (WPs) allow generation of any single photon state and the Bell states
A delay line is inserted to control the temporal superposition of the photons, which are
injected into the integrated device through single mode fibers (SMFs). 
Interference filters determine the photon bandwidth $\Delta\lambda=6nm$.
(ii) Integrated quantum walk circuit realized by ultrafast laser writing technique (see inset on the right).
(iii)Measuring apparatus: the chip output is divided by a beam splitter (BS) and magnified through a set of two lenses. The photons coupled to multimode fibers (MMFs) are then delivered to single photon counting modules. 
The MMFs are mounted on motorized translation stages in order to select an arbitrary combination of two output ports and measure 
two-photon coincidences. Polarization controllers (PC) are used before the chip to compensate polarization rotations induced by the fibers.}
\label{fig:setup}
\end{figure}

In a previous work we have demonstrated that femtosecond laser technology can produce high-quality waveguides able to support polarization entangled 
photon states \cite{sans10prl,SI:chip}. 
In these femtosecond laser written waveguides, birefringence is low and does not affect significantly the coherence of the photons. Anyway the {guided modes for the two polarizations are still slightly different (as shown in Fig. \ref{fig:couplers}a) and this results} in a residual polarization dependence in the properties of the fabricated {\textit{DCs}. In fact, the coupling coefficient depends on the overlap integral between the two {guided} mode profiles and is indeed {quite sensitive to even} small differences in the mode dimensions \cite{szam07oex}}. When several devices are cascaded, as in the case of a \textit{DC} array for implementing a quantum walk, small differences in the splitting ratios would accumulate and in the end affect the indistinguishability of the two polarizations.

The three-dimensional capabilities of the present technology can be exploited to tailor the polarization behaviour of the DCs. We have fabricated several \textit{DC}s with the waveguides lying on planes at different angles $\theta$ with respect to the horizontal (see inset of Fig. \ref{fig:couplers}b), but fixed interaction length and spacing between the waveguides. It can be observed that the ratio between the measured coupling coefficients for the two polarizations ($C_V$ and $C_H$) depends on $\theta$, as shown in Fig. \ref{fig:couplers}b. In particular there exists an angle for which the ratio between the two coefficients 
is one, i.e. {the coupler becomes polarization insensitive.
In order to realize an experimental implementation of a discrete quantum walk with photons, we fabricated a network of \textit{DC}s, all realized with the tilted geometry described {above}, {where the two waveguides are brought at $11 \mu m$ distance, at an angle of $62 ^\circ$, in the interaction region, thus} guaranteeing 
the polarization independence. The length of the interaction region {is chosen as $L = 2.1 mm$} in order to obtain a balanced splitting ratio.

In the interaction region the two waveguides are at different depths in the glass. To connect one coupler to the following, {we designed a structure where the
 waveguides continuously vary the depth, as shown with the color codes in Fig. \ref{fig:couplers}c. The basic cell of the {network}, 
depicted in Fig. \ref{fig:couplers}d, acts as a Mach-Zehnder interferometer. For the correct operation of the quantum walk all the interferometers 
present in the network must be phase balanced. This is intrinsically achieved with the highly symmetric three-dimensional geometry implemented in the network (Fig. \ref{fig:couplers}c and \ref{fig:couplers}d). The two central waveguides of the structure start with 
an initial separation of $250 \mu m$ to couple the device with a single-mode fiber array, while at the output the waveguides are separated by $70 \mu m$. The whole chip is $32 mm$ long.

To carry out and characterize the different quantum walks we adopted the experimental apparatus reported in Fig. \ref{fig:setup}. 
Different single photon and two photon states were injected in the {network of DCs}.
{
The singlet-triplet transition within the Bell basis was performed by applying a $\pi$-shift in the phase $\phi$ of the state $\frac{1}{\sqrt{2}}(\ket{H}_A\ket{V}_B+e^{i\phi}\ket{V}_A\ket{H}_B)$ through rotations of HWP and QWP (see Fig. \ref{fig:setup}). Fine phase adjustment was performed by a voltage liquid crystal (LC) device inserted on mode $k_A$}.

The output of the integrated device is collected by a suitable telescope,  splitted through a bulk beam splitter and then coupled to two multimode fibers (MMFs). By independently translating the MMFs on the arms $C$ and $D$ we
select the output ports to detect, respectively $I$ and $J$, and {measure} the single photon signals $S_C(I)$ and $S_D(J)$ and 
two-photon coincidences $C_{CD}(I,J)$.

As a first measurement we characterized the quantum walk circuit with single photons injected {in either mode $k_A$ or $k_B$. 
By measuring the output signals $S_C(I)$ we obtained the single particle distributions. {
In order to demonstrate the polarization insensitivity of our device, we repeated this measurement by injecting light in different 
polarization states -horizontal, vertical, diagonal and antidiagonal- always observing very similar distributions \cite{SI:prob}. Each experimental distribution was compared with the expected one by the similarity 
$S=(\sum_{i,j}{\sqrt{D_{ij}D_{i,j}^{\prime}}})^2/\sum_{i,j}{D_{ij}}\sum_{i,j}{D_{ij}^{\prime}}$, which is a generalization of the 
classical fidelity between two distributions $D$ and $D^{\prime}$.} The mean value over the tested input polarization states is $S_{1ph}=0.992\pm0.002$.}

As a second step we injected two-photon entangled states.
The distribution of the triplet and singlet states $\ket{\Psi^+}$ ($\phi=0$) and $\ket{\Psi^-}$ ($\phi=\pi$) emerging from the quantum walk was reconstructed by measuring the coincidence counts $C_{CD}(I,J)$ for each combination of the indices $I$ and $J$.
The measured bosonic and fermionic distributions compared with expected ones are reported in Fig. \ref{fig:figure4}a-b 
for $\ket{\Phi^+}$ and $\ket{\Psi^-}$. As done for the measurements on the sigle photon quantum walk, we plotted the probability distributions for the walkers to be in the
final positions $i,j$ of the quantum walk which is
related to the probability of photons to emerge from the output ports $I$ and $J$ of the BS array \cite{SI:prob}.

\begin{figure}
\includegraphics[width=8.5cm]{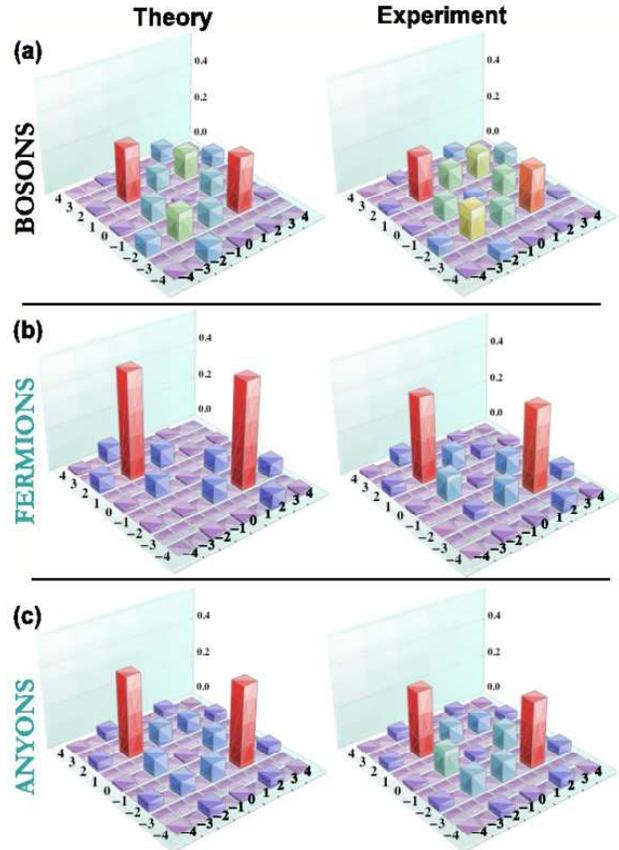}
\caption{Two-particle quantum walks: ideal (left) and measured (right) distributions of (a) bosonic, (b) fermionic and (c) anyonic {(with $\phi=\pi/2$)} quantum walks.}
\label{fig:figure4}
\end{figure}
In {Fig.} \ref{fig:figure4} we observe how the particle symmetry affects the quantum walks.
Note that
some of the diagonal elements of the
fermionic two-particle walk are nonzero both in the theoretical and
experimental distribution. Indeed {Fig.} \ref{fig:figure4} reports the probability
distribution of the walk {positions} and not {of} the physical spatial modes.
{In fact,} the expected probability to measure two fermions
over the same output {spatial} mode is vanishing \cite{SI:prob}.

Furthermore, by considering a generic phase $\phi$ (different from 0 and $\pi$), it is possible to simulate the behaviour of the quantum walk of two anyons, particles with a non semi-integer spin that represent a generalization of fermions and bosons \cite{wilc82prl}. Precisely, the
entangled state $\ket{\Psi^{\phi}}=\frac{1}{\sqrt{2}}(\ket{H}_A\ket{V}_B+e^{i\phi}\ket{V}_A\ket{H}_B)$ simulates two anyons characterized by
creation operators satisfying $c_ic_j=e^{i\phi}c_jc_i$ and $c_ic^\dag_j=e^{i\phi}c^\dag_jc_i+\delta_{ij}$. These systems
{exhibit} both bunching and anti-bunching behaviours (i.e. diagonal and off-diagonal elements in the final distribution). {As a further measurement, we therefore} prepared some anyonic states $\ket{\Psi^{\phi}}$, in particular with $\phi=\frac{\pi}{4},\frac{\pi}{2},\frac{3}{4}\pi$, and measured the output probabilities. In Fig. \ref{fig:figure4}c the distribution for $\phi=\frac{\pi}{2}$ is reported as an example of an anyonic behaviour.

The experimental data can be compared with the theoretical distributions by the similarity 
obtaining $S_{bos}=0.982\pm0.002$ and $S_{fer}=0.973\pm0.002$ for the bosonic and fermionic quantum walk and $S_{any}^{\pi/4}=0.987\pm0.002$, $S_{any}^{\pi/2}=0.988\pm0.001$ and $S_{any}^{3\pi/4}=0.980\pm0.002$ for the anyonic quantum walks with $\phi=\frac{\pi}{4},\frac{\pi}{2},\frac{3}{4}\pi$, respectively. The obtained results are in good agreement with the expected behaviours.

In conclusion, we presented the behavior of a discrete quantum walk based on an integrated array of symmetric, polarization insensitive, directional couplers in which two-photon polarization entangled states are {injected. Exploiting} the different statistics of singlet and triplet entangled states, such scheme allowed us to simulate how symmetric and antisymmetric particles travels through the quantum walk.

The insensitivity to photon polarization, high-accuracy in the phase control and intrinsic scalability of the integrated multi-DC network presented in this work, pave the way to further advanced investigations on  complexity physics phenomena.
For instance, by introducing suitable static and dynamic disorder in the walk it would be possible to simulate the interruption of diffusion in a periodic lattice, like Anderson localization \cite{ande58pr,torm02pra,keat07pra,yin08pra}, and the noise-assisted quantum transport effect {\cite{mohs08jcp,caru11pra}.

During the final revision of this manuscript, the simulation of quantum statistics with entangled photons within a continuous quantum walk has been reported online \cite{matt11arx}.
This work was supported by EU-Project CHISTERA-QUASAR, FIRB-Futuro in
Ricerca HYTEQ and PRIN 2009.

\end{document}


\section*{Supplementary Information}
\setcounter{figure}{4}

\textbf{Chip fabrication and {single photon} characterization} - Waveguides were micromachined in borosilicate glass {substrate (Corning EAGLE2000)} using a cavity-dumped Yb:KYW mode-locked oscillator, which delivers $300 fs$ pulses at $1 MHz$ repetition rate. For the waveguide fabrication pulses with $220 nJ$ energy were focused $170 \mu m$ under the glass surface, using a $0.6 N.A.$ microscope objective while the sample was translated at a constant speed of $40 mm/s$ by high precision, three-axis air-bearing stages (Aerotech FiberGlide 3D).
Measured propagation losses are $0.5 dB/cm$ and coupling losses to single mode fiber arrays at the input facet are about $1 dB$. Birefringence of these waveguides is on the order of $B = 7 \times 10^{-5}$, as characterized in \cite{sans10prl}.

\begin{figure}[ht]
\includegraphics[width=8.6cm]{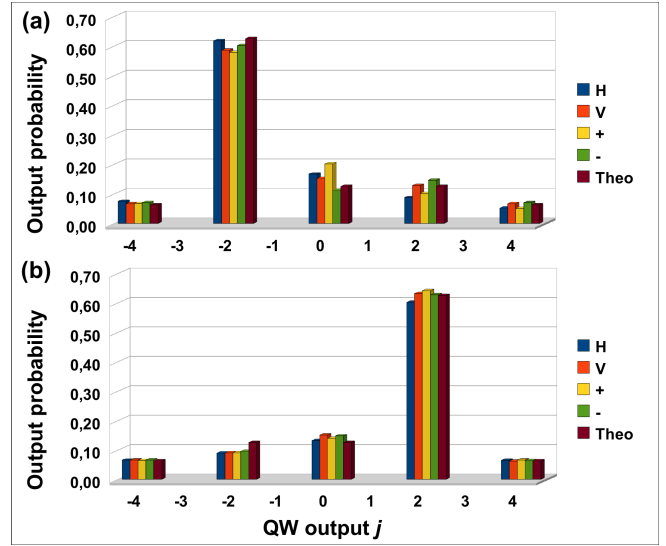}
\caption{Measured {output} probability distributions and {theoretical expectations for} a 4 steps quantum
walk {of a single particle injected into mode $k_A$ (\textit{a}) and mode $k_B$ (\textit{b}).
The experimental results demonstrate that the DCs network leads to the same quantum dynamics independently
from the initial polarization state and in agreement with the theory}.}
\label{fig:figure3}
\end{figure}

In order to evaluate the effect of this residual birefringence we measured the single photon distributions for horizontal, vertical, diagonal and antidiagonal polarized light. The obtained probability distributions for photons injected in both the modes $k_A$ and $k_B$ are reported in Fig. \ref{fig:figure3}. It is clear from the plot that the behaviours are polarization independent, as confirmed by the similarities associated to each distribution, whose values are reported in table \ref{tab:similarity}, both for photons injected in mode $k_A$ and $k_B$.
\begin{table}[htdp]
\begin{center}
\begin{tabular}{|c|c|c|}
\hline
Input state & Similarity $k_A$ & Similarity $k_B$\\\hline\hline
H & $0.991\pm0.002$ & $0.992\pm0.002$\\\hline
V & $0.997\pm0.002$ & $0.993\pm0.002$\\\hline
+ & $0.983\pm0.002$ & $0.991\pm0.002$\\\hline
- & $0.995\pm0.002$ & $0.992\pm0.002$\\\hline
\end{tabular}
\end{center}
\caption{Similarities between the experimental distributions and the theoretical one for horizontal (H), vertical (V), diagonal (+) and antidiagonal (-) polarized input photons in the single-particle regime.}
\label{tab:similarity}
\end{table}

\textbf{Equivalence between entangled states and boson-fermion quantum walk}
\label{sec-walk}.
Let us consider a generic $T$ step {single-photon quantum walk implemented as in Fig. 1b}.
If the particle
is injected at mode $\ket{J}$, with input state $a^\dag_J\ket0$, 
the walk {performs} a unitary transformation on the creation
operator, namely $a^\dag_J\rightarrow \sum_KU_{JK}a^\dag_K$.
Let's now consider two photons injected into the walk in a polarization entangled state 
$\ket{\Psi^\pm_{IJ}}\equiv\frac{1}{\sqrt2}(a^\dag_Ib^\dag_J\pm a^\dag_Jb^\dag_I)\ket0$
where $a^\dag$ and $b^\dag$ are the horizontal and vertical polarization 
creation {operators, respectively}.
The evolution of the walk becomes
\be
\ket{\Psi^\pm_{IJ}}\rightarrow \frac{1}{\sqrt2}\sum_{K,L}
(\underbrace{U_{IK}U_{JL}\pm U_{JK}U_{IL}}_{\psi^\pm_{IJ,KL}})a^\dag_Kb^\dag_L\ket0
\eeq
The probability of detecting one photon at $K$ and the other at $L$
(without {measuring the} polarization) is given by
\beq
p_\pm(I,J;K,L)=
\left\{
\begin{aligned}
&|\psi^\pm_{IJ,KL}|^2&&\text{for }L\neq K
\\
&\frac12|\psi^\pm_{IJ,KL}|^2&&\text{for }L=K
\end{aligned}
\right.
\eeq
\begin{figure*}[ht!!!!]
\includegraphics[width=18cm]{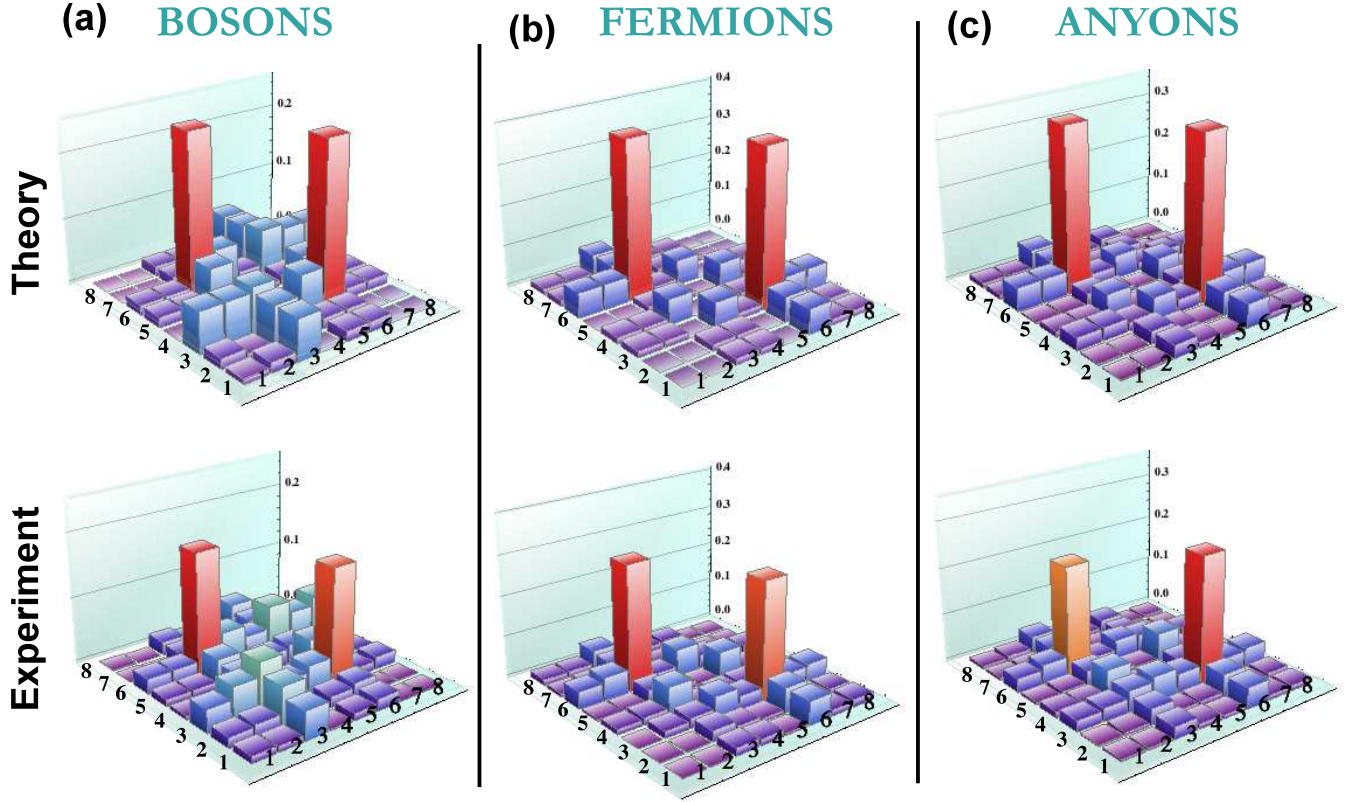}
\caption{Theoretical (top) and measured (bottom) distributions for two-photon states emerging from the integrated array of BS. Depending on the symmetry of the two-photon Bell state injected into the device bosonic (a), fermionic (b) and anionic (c) behaviors are observed.}
\label{fig:modi}
\end{figure*}
It is easy to show that $p_+(I,J;K,L)$ and $p_-(I,J;K,L)$ respectively correspond to the
probabilities of detecting at positions $K$ and $L$ 
two bosons or fermions injected into the modes $I$ ad $J$ of the quantum walk.
In fact, if at the input state we have two identical bosons characterized by commuting creation operator 
$d^\dag_I$ and $d^\dag_J$, the output state is $d^\dag_Id^\dag_J\ket0\rightarrow\frac12\sum_{K,L}
\psi^+_{ij,KL}d^\dag_Kd^\dag_L\ket0
$. The probability of detecting one boson at $K$ and the other at $L$
is precisely $p_+(I,J;K,L)$.
If at the input state we have two identical fermions characterized
 by anticommuting creation operator $c^\dag_I$ we have
$c^\dag_Ic^\dag_J\ket0\rightarrow\frac12\sum_{K,L}
\psi^-_{IJ,KL}c^\dag_Kc^\dag_L\ket0$
and the probability of detecting one fermion at $K$ and the other at  $L$
is $p_-(I,J;K,L)$.
Note that the dynamics of two-particle quantum walk 
cannot be reconstructed by multiplying the output probabilities ($|U_{JK}|^2$)
of two single-particle walks. 

Furthermore it is evident from {Fig. \ref{fig:figure3}} that the plotted probability distributions 
refer to the position
$j$ of the walker and not to the output sites
$J$ of the BS array. Indeed, referring to Fig. 1b {\color{cyan}}, for a walk with $T^*$ steps, the relation between the probabilities of 
photons emerging from one of the $N=2T^*$ outputs of the BS array ($P^{BS}_J$) and the final position of the walker ($p^{walk}_j$) 
is:

\begin{equation}
\begin{aligned}
\label{eq:prob}
&p_{-T^*}^{walk}=P_{1}^{BS},
\\
&p_{-T^*+2k}^{walk}=P_{2k}^{BS}+P_{2k+1}^{BS},\ k=1,...,T^*-1
\\
&p_{T^*}^{walk}=P_{2T^*}^{BS},
\end{aligned}
\end{equation}
Then it is clear that in an array with an even
(odd) number of steps the output ports J are grouped into
the even (odd) final positions $j$ of the walker.

These relations between the output probabilities of the BS array and the sites of the QW
cause the diagonal elements of the two-particle fermionic distribution to be non-vanishing, as clearly shown in Fig. \ref{fig:modi}(b).
This can be ascribed to the relation between the outputs $J$s of the BS array and the positions $j$s of the walker: the symmetrization postulate of quantum mechanics makes the probability of the photons emerging from the same  output of the BS array to be zero -i.e. $P^{BS}_{JJ}=0$-, all the same, according to the relations (\ref{eq:prob}) generalized to the case of two-particle walk, a probability $p_{jj}^{walk}\neq0$ for the walk distribution can be observed.

This is not surprising because two degrees of freedom are involved, the lattice site and the internal coin state: the antisymmetry for fermions in the same lattice site $k$ is provided by the different internal coin state, one particle is in the $\ket{U}$ state and the other one in the $\ket{D}$ state with a total wavefunction of the form:
\begin{equation}
	\ket{\Psi_{kk}}=\frac{1}{\sqrt{2}}\left[\ket{j,U}-\ket{j,D}\right]
\end{equation}
which is clearly antisymmetric.